\documentclass{article}
\usepackage{titlesec}
\usepackage{authblk}
\usepackage{multirow}
\usepackage{setspace}

\usepackage[english]{babel}
\usepackage[letterpaper,top=2cm,bottom=2cm,left=2cm,right=2cm,marginparwidth=1.75cm]{geometry}

\usepackage{amsmath}
\usepackage{graphicx}
\usepackage{float}
\graphicspath{ {./Figures/} }
\usepackage[colorlinks=true, allcolors=blue]{hyperref}

\usepackage{tikz}
\usepackage{pifont}
\usepackage{cleveref}
\usepackage{cite}
\usepackage{array}

\usepackage{listings}
\usepackage{xcolor} 

\definecolor{codegray}{rgb}{0.5,0.5,0.5}
\definecolor{codegreen}{rgb}{0,0.6,0}
\definecolor{backcolour}{rgb}{0.95,0.95,0.92}

\lstdefinestyle{my_code_style}{
    backgroundcolor=\color{backcolour},   
    commentstyle=\color{codegreen},
    morecomment=[l]{\#},
    keywordstyle=\color{magenta},
    numberstyle=\tiny\color{codegray},
    stringstyle=\color{codepurple},
    basicstyle=\footnotesize\ttfamily,
    breakatwhitespace=false,         
    breaklines=true,                 
    captionpos=b,                    
    keepspaces=true,                 
    numbers=left,                    
    numbersep=5pt,                  
    showspaces=false,                
    showstringspaces=false,
    showtabs=false,                  
    tabsize=2
}

\lstdefinelanguage{yaml}{
    keywords={true, false, null},
    keywordstyle=\color{darkgray}\bfseries,
    basicstyle=\footnotesize\ttfamily,
    morecomment=[l]{\#},
    commentstyle=\color{purple}\ttfamily,
    stringstyle=\color{red}\ttfamily,
    moredelim=[l][\color{orange}]{\&},
    moredelim=[l][\color{magenta}]{*},
    morestring=[b]',
    morestring=[b]",
}

\renewenvironment{abstract}{%
    \noindent\begin{center}\bfseries\abstractname\end{center}%
    \quotation
}{%
    \endquotation
}

\title{ChemPlasKin: a general-purpose program for unified gas and plasma kinetics simulations}
\author{{Xiao Shao$^{a,*}$, Deanna A. Lacoste$^{a}$, Hong G. Im$^{a}$ }\\
    {\footnotesize \em $^a$CCRC, King Abdullah University of Science and Technolody (KAUST), Thuwal 23955-6900, Saudi Arabia} \\
    {\footnotesize $^*$E-mail: xiao.shao@kaust.edu.sa}
    }

\begin{document}
\date{} 
\onehalfspacing

\vspace{40pt}
\maketitle
\begin{abstract}

This work introduces ChemPlasKin, a freely accessible solver optimized for zero-dimensional (0D) simulations of chemical kinetics of neutral gas in non-equilibrium plasma environments. By integrating the electron Boltzmann equation solver, CppBOLOS, with the open-source combustion library, Cantera, at the source code level, ChemPlasKin computes time-resolved evolution of species concentration and gas temperature in a unified gas-plasma kinetics framework. The model allows high fidelity predictions of both chemical thermal effects and plasma-induced heating, including fast gas heating and slower vibrational-translational relaxation processes. Additionally, a new heat loss model is developed for nanosecond pulsed discharges, specifically within pin-pin electrode configurations. With its versatility, ChemPlasKin is well-suited for a wide range of applications, from plasma-assisted combustion (PAC) to fuel reforming. In this paper, the reliability, accuracy and efficiency of ChemPlasKin are validated through a number of test problems, demonstrating its utility in advancing gas-plasma kinetic studies.

\vspace{10pt}

\vspace{10pt}
\noindent {\em Source code:} \url{https://github.com/ShaoX96/ChemPlasKin}
\end{abstract}

\vspace{40pt}

\section{Introduction} 

Non-equilibrium plasma has gained an increasing interest within both the combustion and plasma research communities, owing to its potential to enhance combustion characteristics and fuel reforming \cite{ju_plasma_2015, lacoste_flames_2023, wang_catalyst-free_2022}. The integration of plasma actuation into reacting flows, via kinetic, thermal, and hydrodynamic effects, presents a complex multi-physics challenge that requires a synergistic approach combining both experimental and computational methodologies to gain a comprehensive understanding. Due to the formidable computational complexity and expense associated with higher-dimensional models, detailed chemical kinetic analyses of systems that couple neutral gas with plasma have predominantly been confined to zero or one-dimensional simulations. Such 0D simulations, enriched with detailed gas-plasma kinetics, are not only foundational for the development of kinetic mechanisms but are also pivotal in identifying important pathways of plasma energy transfer. This insight is crucial for creating reduced-order plasma models that can be integrated efficiently into computational fluid dynamics (CFD) solvers \cite{castela_direct_2016, barleon_phenomenological_2023}.

A combined kinetic solver requires the determination of species reaction rates for the gas-phase and plasma. For the latter, the electron energy distribution function (EEDF) is crucial for determining reaction rate coefficients of inelastic electron-impact collisions and calculating electron temperature ($T_e$). In non-equilibrium plasma, EEDF deviates from Maxwellian, and needs to be determined by solving the electron Boltzmann equation (EBE). Table \ref{tab:1} summarizes various modeling approaches for computing chemical kinetics in gas-phase coupled with non-equilibrium plasma available in the literature. Broadly, solver development methodologies fall into three categories: tabulation, operator-splitting, and full coupling. Solvers 1-4 employ either the Bolsig+ EBE solver \cite{hagelaar_solving_2005} or the ZDPlasKin plasma kinetics solver \cite{pancheshnyi_computer_2008} to pre-calculate electron-impact reaction rates to build look-up tables or polynomial fits as functions of reduced electric field ($E/N$) or $T_e$, for use with CHEMKIN \cite{kee_chemkin-ii_1989}, the standard in gas-phase chemical kinetics software. While straightforward, tabulation may be subjected to large errors with significant gas composition changes. Solvers 5-13, following Lefkowitz et al. \cite{lefkowitz_species_2015}, integrate ZDPlasKin with chemical kinetics solvers like CHEMKIN or Cantera \cite{goodwin_cantera_2022}, alternating the integration of plasma and neutral gas kinetics with stepwise exchange of species concentration and temperature data. Solver 14 by Cheng et al. \cite{cheng_plasma_2022} is the only approach known to date that directly couples the BOLOS EBE solver \cite{aluque_aluquebolos_2023} with Cantera, enabling unified gas-plasma kinetics simulations, although the details of its code implementation were not elaborated.

While CHEMKIN and Cantera are well-established open-source modules for combustion community, and ZDPlasKin is for the plasma community, the integration of the two schools for a unified version of gas-plasma kinetics solver is not straightforward, as evidenced by the fact that most of the tools shown in Table \ref{tab:1} are in-house codes and lack open availability. To address this deficiency, we introduce ChemPlasKin, an open-source code optimized for simulating neutral gas-phase chemical reactions in conjunction with non-thermal plasma chemistry within a unified code module framework. Similar to the CERFACS code \cite{cheng_plasma_2022}, ChemPlasKin is developed with a particular interest in for nanosecond repetitively pulsed (NRP) plasma applications, which exhibits a large spectrum of time scales, offering new capabilities for plasma-assisted ignition (PAI) and fuel reforming applications.

The main content of this paper is divided into two sections. Section \ref{methodology} outlines the methodologies, including the governing equations, the heat loss model, and code development. Section \ref{validation} presents extensive code validations that demonstrate the consistency and efficiency of ChemPlasKin in comparison with results from the literature. It also highlights the solver's capability in predicting both fast and slow gas heating processes, and assesses the impact of the newly introduced heat loss model.

\newcounter{rownum}
\setcounter{rownum}{0}
\newcommand{\rownumber}{\stepcounter{rownum}\therownum}

\begin{table} [H]
\caption{\label{tab:1} Literature review of publications involving in-house 0D gas-plasma kinetics solvers}
\centering
\begin{tabular}{ll p{4.0cm} p{7cm} ll}
\hline
\textbf{$\#$}   & \textbf{Origin}   & \textbf{Method}  & \textbf{Topic}   & \textbf{Discharge}    & \textbf{Year} \\
\hline
\rownumber & Ohio State  & Bolsig+ precalculation + ChemKin-Pro  & Air/$\mathrm{H_2}$ and air/hydrocarbon kinetics \cite{adamovich_kinetic_2015} & NRP &2015 \\
\rownumber & UC Berkeley  & ZDPlasKin precalculation + CHEMKIN  & $\mathrm{CH_4}$/air ignition \cite{defilippo_modeling_2016} & NRP &2016\\

\rownumber & UT Austin & Bolsig+ precalculation + CHEMKIN & $\mathrm{CH_4}$/air and $\mathrm{C_2H_4}$/air ignition \cite{deak_plasma-assisted_2021} & NRP &2021\\

\rownumber & TU/e & Bolsig+ precalculation + Sundials IDA & $\mathrm{CH_4}$ and $\mathrm{H_2}$ oxidation in Ar \cite{hazenberg_chemical_2023} & DBD &2023\\

\rownumber & Princeton & ZDPlasKin + CHEMKIN & $\mathrm{CH_4/O_2/He}$ oxidization \cite{lefkowitz_species_2015} & NRP &2015\\
& & & $\mathrm{CH_4}$/air ignition \cite{mao_effects_2018} & NRP &2018\\
& & & $\mathrm{H_2/O_2/He}$ ignition \cite{mao_effects_2019} & NRP/DC &2019\\
& & & $\mathrm{CH_4/O_2/He}$ ignition \cite{mao_numerical_2019} & NRP/DC &2019\\
& & & $\mathrm{H_2}$/air ignition \cite{wang_numerical_2020-1} & NRP &2020\\
& & & N-dodecane/$\mathrm{O_2/N_2}$ kinetics \cite{zhong_kinetic_2021} & NRP &2021 \\
& & & $\mathrm{NH_3/O_2/He}$ ignition \cite{faingold_numerical_2021} & NRP &2021\\
& & & $\mathrm{NH_3}$/air ignition \cite{faingold_plasma_2022} & NRP &2022\\
& & & $\mathrm{N_2O/NO_x}$ mechanism in $\mathrm{NH_3}$ oxidation  \cite{zhong_understanding_2023} & DBD & 2023 \\
& & & $\mathrm{NH_3}$/air ignition \cite{mao_ignition_2024} & NRP &2024\\
& & & n-pentane/air oxidation \cite{liu_kinetic_2024} & NRP &2024\\

\rownumber & USC & ZDPlasKin + Cantera & DME/$\mathrm{O_2/Ar}$ and $\mathrm{C_3H_8/O_2/Ar}$ ignition \cite{gururajan_transient_2017} & nanosecond &2017\\

\rownumber & Tsinghua & ZDPlasKin + CHEMKIN & DME oxidation kinetics \cite{zhang_exploring_2021} & DBD & 2021 \\

\rownumber & UMN & ZDPlasKin + CHEMKIN & $\mathrm{NH_3}$ pyrolysis and combustion \cite{taneja_comparing_2021} & NRP &2021\\
& & & $\mathrm{NH_3}$/air ignition and $\mathrm{NO_x}$ emission \cite{taneja_nanosecond_2022}  & NRP &2022 \\
& & & Plasma-based global pathway analysis \cite{johnson_plasma-based_2023} & NRP &2023 \\

\rownumber & KAUST & ZDPlasKin + CHEMKIN & Lean $\mathrm{H_2/O_2}$ kinetics \cite{snoeckx_kinetic_2022} & DBD &2022\\
& & & Rich $\mathrm{H_2/O_2}$ kinetics \cite{snoeckx_kinetic_2023} & DBD &2022\\
& & & $\mathrm{NH_3}$ cracking \cite{bang_kinetic_2023} & DBD &2023\\
& & & $\mathrm{O_3}$ kinetics \cite{bang_temperature-dependent_2023} & DBD &2023\\

\rownumber & Birmingham & ZDPlasKin + Cantera & $\mathrm{NH_3/N_2/O_2/He}$ combustion \cite{shahsavari_nanosecond_2022} & NRP &2022\\

\rownumber & MIT & ZDPlasKin + Cantera & 1D $\mathrm{CH_4/air}$ flame \cite{pavan_modeling_2023} & NRP-DBD &2023\\

\rownumber & XJTU & ZDPlasKin + in-house chemistry solver & $\mathrm{NH_3/N_2/O_2}$ ignition \cite{qiu_numerical_2023} & NRP/DC &2023\\

\rownumber & SDU & ZDPlasKin + CHEMKIN & $\mathrm{CH_4/O_2/He}$ ignition \cite{xin_numerical_2024} & NRP-SDBD &2024\\

\rownumber & CERFACS & BOLOS + Cantera & $\mathrm{CH_4}$/air mechanism \cite{cheng_plasma_2022} & NRP &2022\\
& & & $\mathrm{CH_4}$/air phenomenological model \cite{barleon_phenomenological_2023} & NRP &2023 \\

\hline
\end{tabular}
\end{table}

\section{Methodology}\label{methodology}
\subsection{Governing equations}

The governing equations build upon the previous work by Cheng et al. \cite{cheng_plasma_2022} and are presented in a more generalized format. ChemPlasKin integrates the mass fraction $Y_k$ and gas temperature $T_{\text{gas}}$ over time $t$ in a 0D neutral gas-plasma reactor for a total of $N$ species:
\begin{equation}
    \label{eq:Y_k}
    \frac{dY_k}{dt} =\frac{W_k}{\rho} \dot{\omega}_k,
\end{equation}
\begin{subequations}
    \begin{align}
        \label{eq:Tg_cv}
        \rho c_v \frac{dT_{\text{gas}}}{dt} &= - \sum_{k=1}^N \dot{\omega}_k u_k W_k + \dot{E}^p + \sum_{k}{(-\dot{E}_{\text{vib}}^k + \dot{R}_{\text{VT}}^k)}, \quad \text{(constant volume)} \\
        \label{eq:Tg_cp}
        \rho c_p \frac{dT_{\text{gas}}}{dt} &= - \sum_{k=1}^N \dot{\omega}_k h_k W_k + \dot{E}^p + \sum_{k}{(-\dot{E}_{\text{vib}}^k + \dot{R}_{\text{VT}}^k)}. \quad \text{(constant pressure)}
    \end{align}
\end{subequations}
Here, $W_k$ and $\dot{\omega}_k$ denote the molecular weight and molar production rate of species $k$, respectively. $\rho$ represents the density. The terms $c_v$ and $c_p$ signify the mass heat capacities at constant volume and constant pressure, respectively. The term $u_k$ refers to the internal energy of species $k$, and $h_k$ represents the enthalpy of species $k$.
The total plasma energy $\dot{E}^p$ is derived from non-elastic electron collision processes $\mathcal{P}$: 
\begin{equation}
    \label{eq:Ep}
    \dot{E}^p = \sum_{j \in \mathcal{P}} \varepsilon_\text{th}^j q_j,
\end{equation}
where $\varepsilon_\text{th}^j$ and $q_j$ are the threshold energy and net molar production rate for each process $j$, respectively.

A significant portion of plasma energy, not directly contributing to fast gas heating, is stored in the vibrational states of species $k$, denoted by $\dot{E}_\text{vib}^k$. This energy is released to facilitate slow gas heating, primarily through vibrational-translational (V-T) relaxation, at a rate $\dot{R}_\text{VT}^k$. Given the system's non-equilibrium nature, a dedicated equation for the vibrational energy $e_\text{vib}^k$ is necessary:
\begin{equation}
    \label{eq:e_vib}
    \frac{d e_{\text{vib}}^k}{dt} = \dot{E}_\text{vib}^k - \dot{R}_\text{VT}^k.
\end{equation}
$\dot{R}_\text{VT}^k$ is modeled using the V-T relaxation timescale $\tau_\text{VT}^k$:
\begin{equation}
    \label{eq:R_VT}
    \dot{R}_\text{VT}^k = \frac{e_{\text{vib}^k}}{\tau_\text{VT}^k}.  
\end{equation}
Extending the two-component mixture relaxation formula in \cite{millikan_systematics_1963} to multiple species, we obtain:
\begin{equation}
    \label{eq:tau_VT}
        \left(\tau_\text{VT}^k\right)^{-1} = \sum_m \left( X_m/\tau_\text{VT}^{k,m} \right),  
\end{equation}
where $X_m$ represents the mole fraction of collider $m$ and $\tau_\text{VT}^{k,m}$ the relaxation time for oscillator $k$ in a high dilution of $m$. The condition $m=k$ indicates relaxation in pure gas $k$. Experimental data fitting lines in \cite{millikan_systematics_1963} are expressed as:
\begin{equation}
    \label{eq:tau_p}
        p_0 \tau_\text{VT}^{k,m} = \text{exp}\left(a(T^{-1/3} - b)-18.42 \right), \ \mathrm{[atm \cdot s]},
\end{equation}
with $p_0$ denoting total pressure in atm, and $(a,b)$ representing the fitting parameter pair for a $k\text{-}m$ mixture. Specifically, for air-dominated systems where vibrational energy is primarily stored in $\mathrm{N_2}(v)$ for $E/N > 50$ Td, it is pertinent to consider $\mathrm{N_2, O_2,}$ and $\mathrm{O}$ as major collision partners. For $m = \mathrm{N_2}$, $(a,b)=(220, 0.03)$ are directly obtained from \cite{millikan_systematics_1963}. For $m = \mathrm{O_2}$, $(a,b)=(162, 0.03)$ are inferred from experimental results at 2500 K \cite{jcolgan_vibrational_1967}. The most efficient V-T relaxation pathway involves reactions with $\mathrm{O(^3P)}$ atoms, characterized by a rate constant of $4.5 \times 10^{-15} (T/300)^{2.1} \ [\mathrm{cm^3/s]}$ \cite{popov_fast_2011}, setting $\tau_\text{VT}^\mathrm{{N_2,O}}$ to $488.5/(p_0 T^{1.1}X_\mathrm{O} )$. Note that these $\tau_\text{VT}^{k,m}$ parameters, are different from those in the phenomenological model by Castela et al.\cite{castela_direct_2016}, underscoring the need for validation and refinement for more precise V-T relaxation predictions.

The model's approach to slow heating via V-T relaxation, encapsulated in equations (\ref{eq:Tg_cv}), (\ref{eq:Tg_cp}), (\ref{eq:e_vib}) and (\ref{eq:R_VT}), consolidates all vibrational states of species $k$ into a singular variable, $e_\text{vib}^k$. This aggregation significantly simplifies the species count, facilitating practicality in multi-dimensional simulations. However, users have the option to model each vibrational state explicitly along with their specific V-T relaxation reactions. In such scenarios, the slow gas heating term $\sum_{k}{(-\dot{E}_{\text{vib}}^k + \dot{R}_{\text{VT}}^k)}$ in equation (\ref{eq:Tg_cv}) and (\ref{eq:Tg_cp}) is unnecessary.
Note that the operator-splitting solvers (5-13) listed in Table \ref{tab:1} utilize a model proposed by Flitti \& Pancheshnyi \cite{flitti_gas_2009} to describe gas heating. In this model, external power deposited into the system is distributed among gas heating, electron energy, and chemical energy. Similarly, ChemPlasKin incorporates this model option for gas temperature calculations, analogous to the formulas used in Lefkowitz et al. \cite{lefkowitz_species_2015}. For constant volume configurations, the governing equation reads:
\begin{equation}
\label{eq:Tg_3}
    \rho c_v \frac{dT_{\text{gas}}}{dt} = - \sum_{k=1}^N \dot{\omega}_k u_k W_k + e N_e v_e E - \frac{3}{2} R T_e \dot{\omega}_e,
\end{equation}
where $e$ represents the elementary charge, $N_e$ the electron number density, $v_e$ the electron drift velocity under the electric field $E$, $R$ the gas constant, and $T_e$ the electron temperature. Employing this model requires the inclusion of reactions related to all vibrational states of species $k$, which significantly increases the number of equations that must be solved.

Accurately depicting a weakly ionized plasma necessitates accounting for the electron distribution function, $f_e$, which adheres to the Boltzmann equation in a six-dimensional phase space \cite{hagelaar_solving_2005}:
\begin{equation}
    \label{eq:f_e}
    \frac{\partial f_e}{\partial t} + \mathbf{v}_e \cdot \nabla f_e - \frac{e}{m} \mathbf{E} \cdot \nabla_v f_e = \left( \frac{\partial f_e}{\partial t} \right)_\text{coll},
\end{equation}
where $\mathbf{v}_e$ represents electron velocity, $e$ the elementary charge, $m$ the electron mass, $\mathbf{E}$ the electric field, and $\nabla_v$ the velocity gradient operator. The term on the right-hand side quantifies $f_e$'s collision-induced rate of change. Employing the widely used two-term approximation allows for the decomposition of equation (\ref{eq:f_e}), with components of $f_e$ renormalized as probability distribution functions. The isotropic part, denoted $F_0$, serves as the electron energy EEDF \cite{hagelaar_solving_2005, alves_foundations_2018}. Reaction rate coefficients for collisions $\mathcal{P}$ are derived from $F_0$ and cross-section $\sigma_j$ by integrating over electron energy $\epsilon$:
\begin{equation}
    \label{eq:k_j}
    k_j = \sqrt{2e/m} \int_0^\infty {\epsilon \sigma_j F_0} d \epsilon, \quad j \in \mathcal{P}.
\end{equation}
Furthermore, $F_0$ facilitates the calculation of electron temperature:
\begin{equation}
    \label{eq:T_e}
    T_e = \frac{2}{3} \int_0^\infty {\epsilon^{3/2} F_0} d \epsilon,
\end{equation}
which is commonly featured in the rate coefficient expressions for recombination reactions involving electrons.

\subsection{Heat loss model for pin-pin NRP discharges}
NRP discharges produced by pin-pin electrodes typically result in ultrafast gas heating within an approximately constant volume, generating a weak shock wave and thermal energy loss due to gas expansion \cite{xu_schlieren_2014, tholin_simulation_2013}. Such hydrodynamic phenomena cannot be directly resolved within the limitations of an 0D framework, leading to potential overestimations of temperature increase if heat loss mechanisms are neglected. This discrepancy may remain negligible for scenarios involving low plasma power; however, it becomes increasingly significant with higher energy deposition and a greater number of nanosecond pulses. To enhance ChemPlasKin's ability to accurately model temperature dynamics, particularly for applications investigating plasma-assisted ignition delay times, we propose the following heat loss model.

 During the nanosecond pulse, the gas within the discharge kernel undergoes a rise in temperature and pressure under constant volume conditions. Once the plasma energy input ceases, the discharge kernel experiences an isentropic expansion from state 1 to state 2, described by:
\begin{equation}
    \label{eq:isentropic}
    \frac{T_2}{T_1} = \left( \frac{\rho_2}{\rho_1} \right) ^{\gamma-1} = \left( \frac{p_2}{p_1} \right) ^{\frac{\gamma-1}{\gamma}},
\end{equation}
where $\gamma$ represents the heat capacity ratio. The propagation of the resulting weak shock wave, occurring on the microsecond timescale, rapidly equalizes the kernel pressure back to the ambient pressure $p_2$. For practical code implementation, the temperature and density at the pulse's conclusion are adjusted to reflect $T_2$ and $\rho_2$ values within a single timestep, for given initial ($p_1$) and final ($p_2$) pressures. This model effectively modifies the gas temperature equation between constant volume and constant pressure conditions to mimic the isentropic process, a flexibility not typically offered by previous solvers listed in Table \ref{tab:1}. 

The heat loss model is augmented to account for radial thermal conduction from the hot discharge kernel to the ambient environment, described by:
\begin{equation}
    \label{eq:q_dot}
    \dot{q}_{loss} = -\lambda \frac{dT}{dr} \frac{A_{\text{dis}}}{V_{\text{dis}}} \approx -\lambda C_0 \frac{T-T_0}{r_{\text{dis}}^2},
\end{equation}
where $\dot{q}_{loss}$ represents the heat loss in $\mathrm{W/m^3}$, $\lambda$ the thermal conductivity, and $r_{\text{dis}}$ the effective radius of the cylindrical discharge characterized by surface area $A_{\text{dis}}$ and volume $V_{\text{dis}}$. Here, $C_0$ is a dimensionless empirical coefficient, typically set to 1.0, and $T_0$ denotes the ambient temperature. To simulate gas temperature dynamics during intervals without plasma energy input, the following equation is actually solved:
\begin{equation}
    \label{eq:Tg_heat_loss}
\rho c_p \frac{dT_{\text{gas}}}{dt} = - \sum_{k=1}^N \dot{\omega}_k h_k W_k + \sum_{k} \dot{R}_{\text{VT}}^k - \lambda C_0 \frac{T-T_0}{r_{\text{dis}}^2}.
\end{equation}

\subsection{Code development}
\subsubsection{Electron Boltzmann equation (EBE) solver}
To enhance performance in simulations, ChemPlasKin incorporates an EBE solver, named CppBOLOS, directly integrated into the Cantera library at the source level. This integration represents a central feature of ChemPlasKin's functionality. Among the available EBE solvers within the plasma community, none provides an open-source C++ implementation that employs the efficient two-term expansion method akin to Bolsig+ \cite{hagelaar_solving_2005}. For instance, while MultiBolt \cite{stephens_multi-term_2018} offers a C++ open-source version (v3.x), its multi-term model tends to be computationally intensive and not essential for a gas-plasma kinetics solver. Addressing this need, we developed a C++ version of BOLOS \cite{aluque_aluquebolos_2023}, an open-source Python solver that follows the algorithms specified in \cite{hagelaar_solving_2005}. Now referred to as CppBOLOS, this solver's development process was significantly assisted by GPT-4 \cite{noauthor_gpt-4_nodate}. It is seamlessly integrated with Cantera, enabling dynamic solutions of the electron Boltzmann equation and real-time updates of the EEDF based on the temporal evolution of gas temperature, mixture composition, and reduced electric field. CppBOLOS accepts input cross-section data in the popular LXCat format \cite{pancheshnyi_lxcat_2012}, which ensures broad compatibility and simplifies data integration. The validation of CppBOLOS is briefly presented in Section \ref{cppbolos}.

\subsubsection{Code architecture}

\begin{figure}[htbp]
\centering
\includegraphics[width=0.75\linewidth]{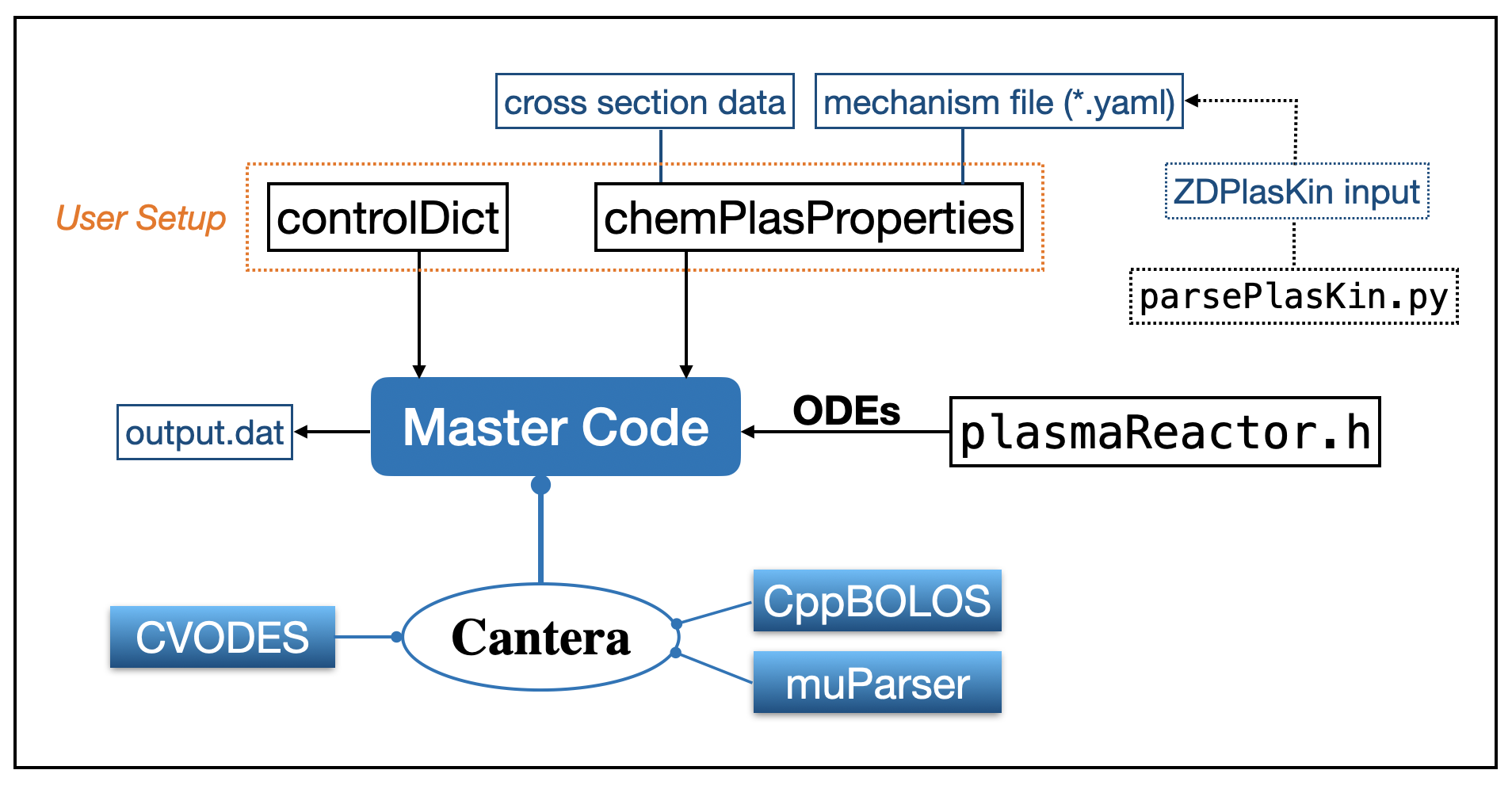}
\caption{\label{fig:code} Code architecture of ChemPlasKin. }
\end{figure}

Figure \ref{fig:code} illustrates the code architecture of ChemPlasKin, highlighting its streamlined approach to handling both plasma and neutral gas kinetics. Key to its design is the use of a unified YAML-type input file for all thermal properties and reactions. This choice facilitates ease of use and integration with existing datasets. ChemPlasKin extends Cantera's reaction module to register inelastic electron-impact reactions classified as type of `Boltzmann', as shown in the examples of Listing \ref{lst:yaml_example_1}:
\begin{lstlisting}[caption=Examples of YAML configurations for Boltzmann type reactions. Vibrational states of $\mathrm{N_2}$ are agrragated and the vibrational energy is stored in \texttt{evib\_N2}., label=lst:yaml_example_1]
# Example 1: Vibrational excitation
- equation: Electron + N2 => Electron + N2  
  type: Boltzmann
  process: N2 -> N2(v1)
  duplicate: true
  energy_transfer: {evib_N2: 0.291_eV}

# Example 2: Electronic excitation
- equation: Electron + O2 => Electron + O2(a1) 
  type: Boltzmann
  process: O2 -> O2(a1)
  energy_transfer: {e_th: 0.977_eV}
\end{lstlisting}

To accommodate the diverse range of rate constant expressions encountered in plasma kinetics, ChemPlasKin employs muParser library \cite{noauthor_muparser_nodate} for parsing any complex mathematical expressions not natively supported by Cantera. Listing \ref{lst:yaml_example_2} shows an example of $\mathrm{e^- + N_4^+}$ recombination reaction, where the reaction rate is calculated as a function of electron temperature. The \texttt{energy\_transfer} entry values enforce zero gas heating and the corresponding energy (3.8 eV) is stored in $e_{\text{vib}}^{\mathrm{N_2}}$.
\begin{lstlisting}[caption=Example of YAML configurations for PlasmaCustomExpr type reactions., label=lst:yaml_example_2]
# Example 3: Recombination
- equation: Electron + N4+ => N2 + N2(C3)
  type: PlasmaCustomExpr
  energy_transfer: {e_th: -3.8_eV, evib_N2: 3.8_eV} # No gas heating
  rateExpr: {A: 2.086e+19, Expr: Te^-0.5}
\end{lstlisting}

For added convenience, an optional parser tool, \verb|parsePlasKin|, is available to convert ZDPlaKin input mechanism files into the human-readable YAML format automatically. This feature is particularly beneficial given the prevalence of ZDPlaKin mechanisms published as supplementary materials in the field. Configuration parameters are set through the text files \verb|controlDict| and \verb|chemPlasProperties|, ensuring straightforward operation.

The core of ChemPlasKin's simulation capabilities is encapsulated in \verb|plasmaReactor.h|, where a single ordinary differential equation (ODE) system—encompassing equations (\ref{eq:Y_k}, \ref{eq:Tg_cv}, \ref{eq:Tg_cp} \ref{eq:e_vib})—is constructed and integrated within the main time loop of the master code. The ODEs are solved using the CVODE solver from the SUNDIALS suite \cite{hindmarsh_sundials_2005, gardner_enabling_2022}, which is efficiently utilized by Cantera for high-precision integration.

\subsubsection{Code efficiency}
ChemPlasKin's unified ODE system offers distinct advantages over operator-splitting techniques, by eliminating splitting errors and enabling much larger timesteps in the main time loop. Although it is suggested that splitting the plasma kinetics could expedite integration due to the higher stiffness during nanosecond pulses, such benefits become marginal during the much longer pulse intervals where plasma stiffness diminishes. Moreover, the plasma kinetics component of ChemPlasKin, akin to that solved by ZDPlasKin, encompasses a broad spectrum of timescales beyond just ultrafast electron-impact reactions. Last but not least, switching between ZDPlasKin and CHEMKIN/Cantera requires reinitialization of their respective ODE solvers at every step, adding extra computational overhead to the solvers in the initial transient to find suitable integration timesteps. Consequently, ChemPlasKin's unified approach not only simplifies the integration process but also enhances overall simulation efficiency compared to the operator-splitting method. To demonstrate this, we have also built a ZDPlasKin-Cantera solver consistent with previous counterparts and compared it with ChemPlasKin, as detailed in Section \ref{h2o2he}.

\section{Code Validation} \label{validation}

This section begins with a brief validation of CppBOLOS against Bolsig+, establishing the groundwork for subsequent evaluations. The subsequent five subsections are carefully structured to assess ChemPlasKin's performance across diverse simulation scenarios. We present comparisons of simulation results between literature sources, our equivalent Cantera-ZDPlasKin solver, and ChemPlasKin in Section \ref{h2o2he}. The capability of ChemPlasKin to predict ultrafast gas heating and radical species production is assessed in Section \ref{spark_air}. The slow gas heating model is validated through experimental data in Section \ref{v-t}, while a practical application case of fuel reforming is examined in Section \ref{dbd}. Finally, the efficacy of the heat loss model is tested in Section \ref{heat_loss}. For the simulations discussed in Sections \ref{spark_air}, \ref{v-t}, and \ref{heat_loss}, we utilize a rigorously validated detailed PAC mechanism involving 100 species and 964 reactions for a methane-air mixture, as developed by \cite{cheng_plasma_2022}.

\subsection{CppBOLOS} \label{cppbolos}
\begin{figure}[H]
\centering
\includegraphics[width=0.6\linewidth]{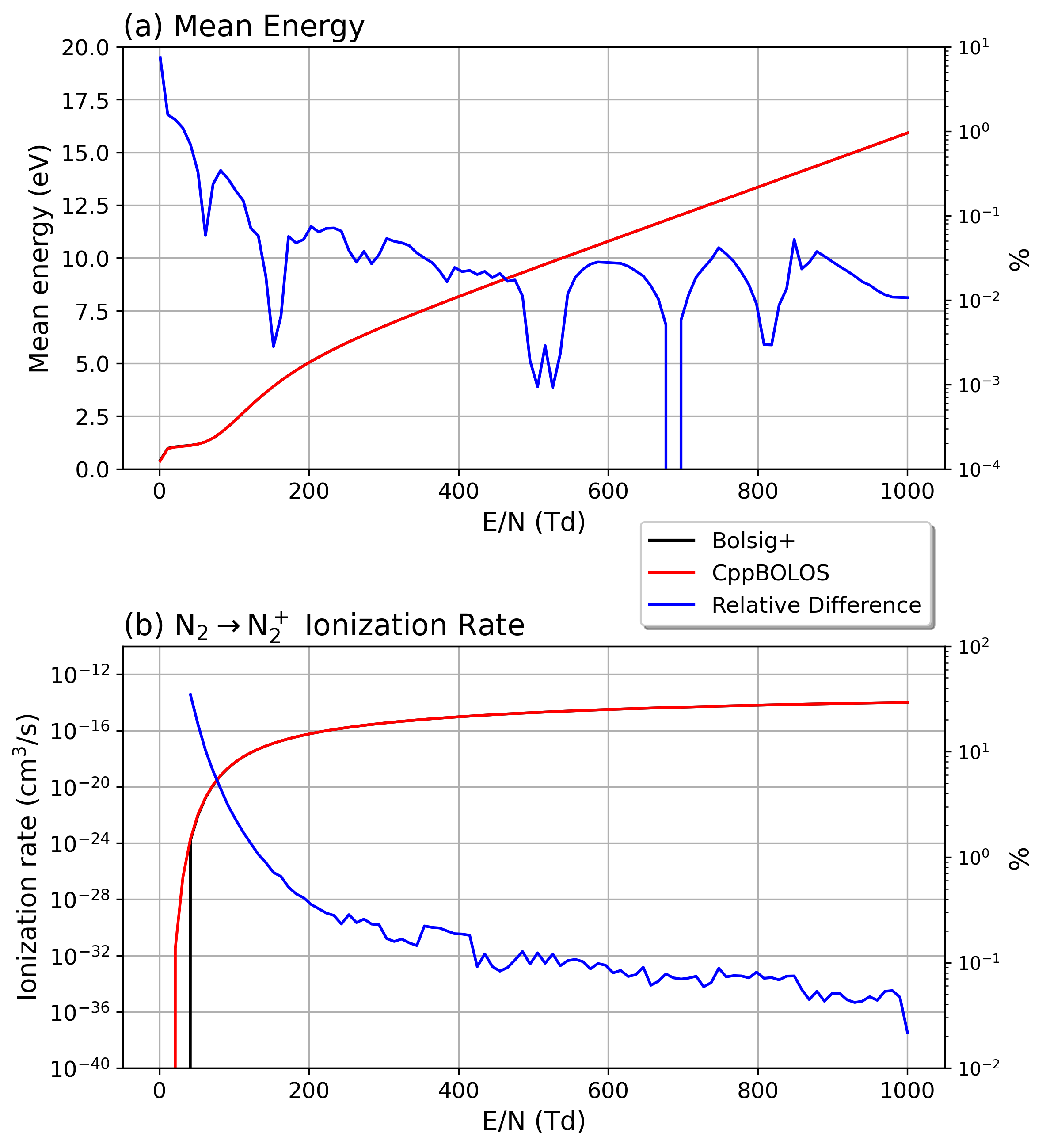}
\caption{\label{fig:cppbolos} Comparison between CppBOLOS and Bolsig+ for pure $\mathrm{N_2}$ at 300 K, using cross-section data from Phelps \cite{pancheshnyi_lxcat_2012}.}
\end{figure}

Figure \ref{fig:cppbolos} illustrates the comparison between CppBOLOS and Bolsig+ for mean electron energy and ionization rate in pure $\mathrm{N_2}$ across a wide range of reduced electric fields ($E/N$). Additionally, relative differences compared to Bolsig+ are plotted, demonstrating satisfactory consistency except at very low $E/N$ values. The large relative differences in ionization rate at low $E/N$ are considered acceptable because the infinitesimally small absolute values have a negligible impact on the accurate prediction of the plasma kinetics system's evolution.

\subsection{Plasma assisted \texorpdfstring{$\mathbf{H_2/O_2/He}$}{H2/O2/He} ignition} \label{h2o2he}
Mao et al. \cite{mao_effects_2019} conducted numerical simulations of $\mathrm{H_2/O_2/He}$ ignition assisted by hybrid NRP and DC discharges using the ZDPlasKin-CHEMKIN solver developed by Lefkowitz et al. \cite{lefkowitz_species_2015}. As no openly available PAC solver exists for use as a benchmark, we integrated ZDPlasKin with Cantera to evaluate the computational efficiency of the popular operator-splitting technique, adhering to the strategy described in \cite{lefkowitz_species_2015}. The Cantera-ZDPlasKin solver includes a Python wrapper of ZDPlasKin that facilitates efficient data communications between the two codes. The timestep in the main time loop is dynamically adjusted to accommodate the nanosecond timescale and exponential growth of electrons in NRP discharges, then gradually relaxed to $10^{-7}$ s during the pulse intervals, balancing accuracy with the computational cost of the operator-splitting method.

The mixture has an initial composition of $\mathrm{H_2:O_2:He=0.1667:0.0883:0.75}$, and the adiabatic system is maintained at atmospheric pressure. The reduced electric field ($E/N$) for NRP discharges is set at 100 Td (1 Td = $\mathrm{10^{-21} V m^2}$), with a frequency of 30 kHz. DC discharges at an $E/N$ of 20 Td are applied between the nanosecond pulses. The deposited plasma energy is fixed at 0.1 $\mathrm{mJ/cm^3}$ per pulse. To ensure consistency, ChemPlasKin operates with the Flitti \& Pancheshnyi \cite{flitti_gas_2009} model for gas temperature. The plasma kinetic mechanism used by \cite{mao_effects_2019}, originally formatted for ZDPlasKin, has been automatically converted into YAML format compatible with ChemPlasKin.

\begin{figure}[H]
\centering
\includegraphics[width=0.6\linewidth]{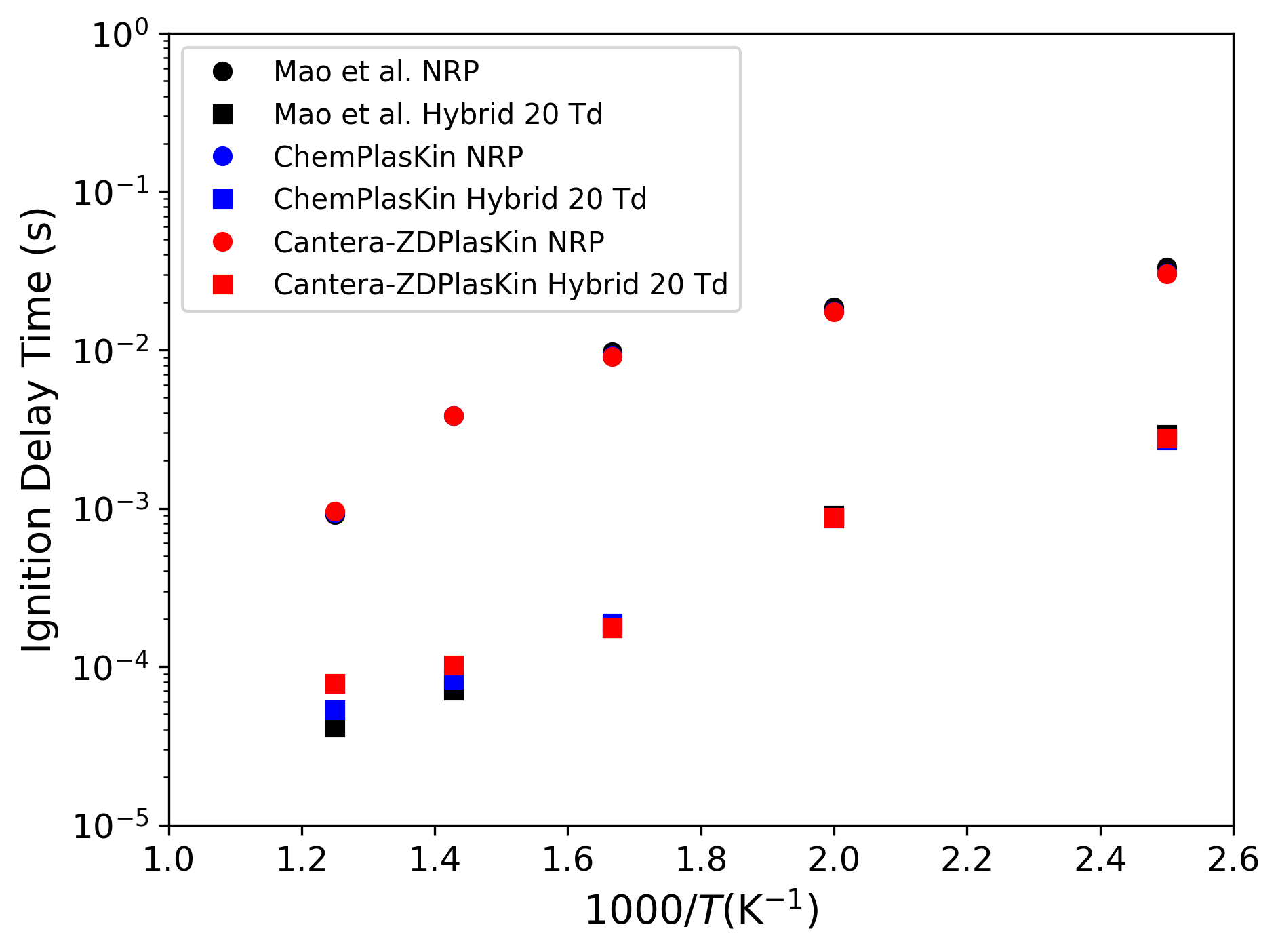}
\caption{\label{fig:IDT} Ignition delay times (IDT) for NRP discharge assisted ignition and NRP/DC hybrid discharge assisted ignition. Results from ChemPlasKin are compared against those from Mao et al. \cite{mao_effects_2019} and our Cantera-ZDPlasKin solver.}
\end{figure}

Figure \ref{fig:IDT} compares these three solvers for their predictions of ignition delay times (IDT) for NRP and NRP/DC hybrid discharges assisted ignition across various initial temperatures. IDT is defined as the point of maximum temperature gradient during the pulse intervals in our simulations. Good agreement is achieved among the three solvers for both discharge types.

\begin{figure}[H]
\centering
\includegraphics[width=0.6\linewidth]{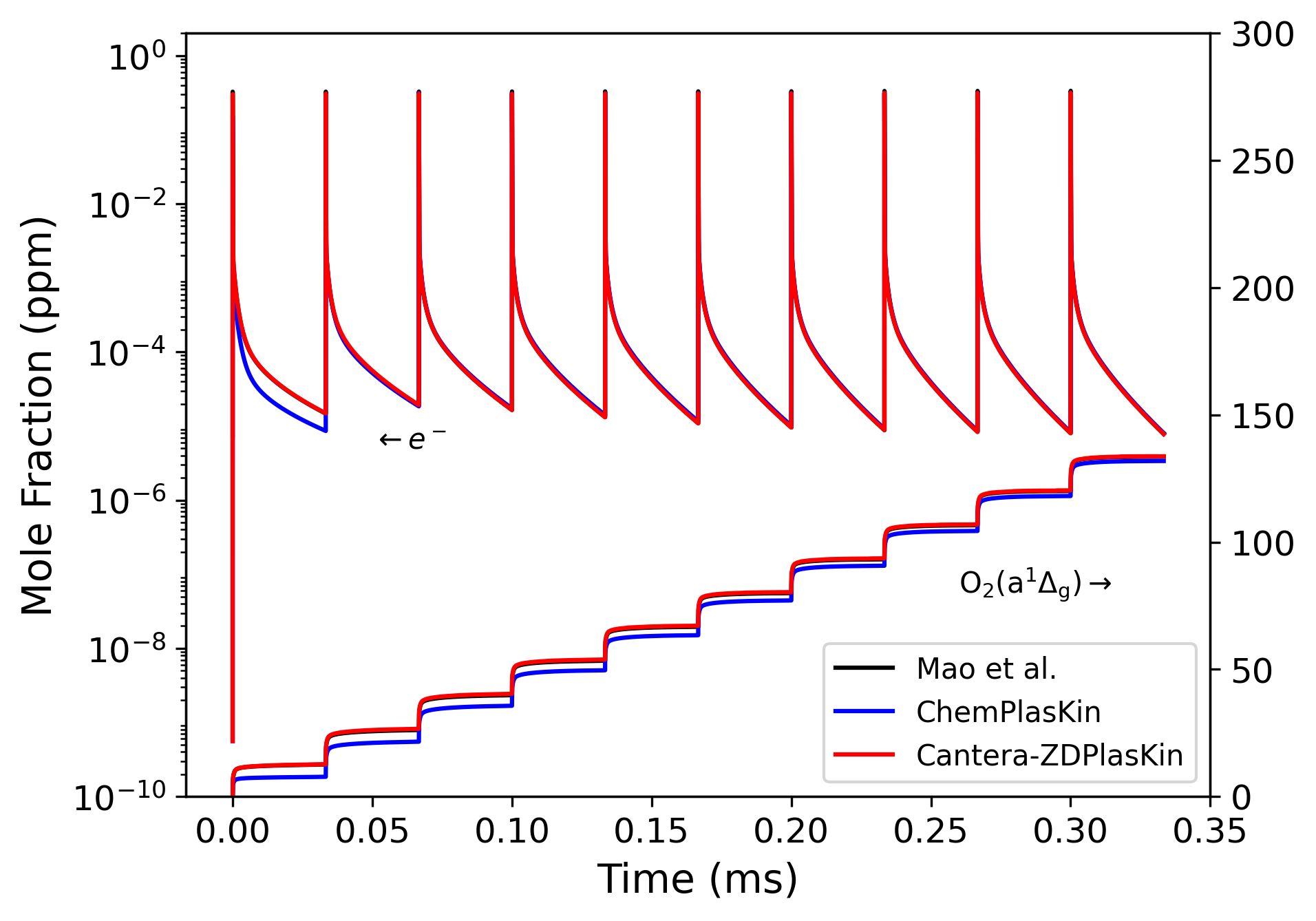}
\caption{\label{fig:O2(1a)_E} Temporal evolution of electron and $\mathrm{O_2(a^1\Delta_g)}$ mole fractions during the first ten pulses of a hybrid discharge at 400 K.}
\end{figure}

The predictive capabilities of the solvers for electron and $\mathrm{O_2(a^1\Delta_g)}$ production over the first ten pulses in the hybrid NRP/DC discharge scenario are displayed in Figure \ref{fig:O2(1a)_E}. These results underscore the performance consistency of our solvers, particularly Cantera-ZDPlasKin, with those reported by Mao et al. \cite{mao_effects_2019}.

\begin{figure}[H]
\centering
\includegraphics[width=0.6\linewidth]{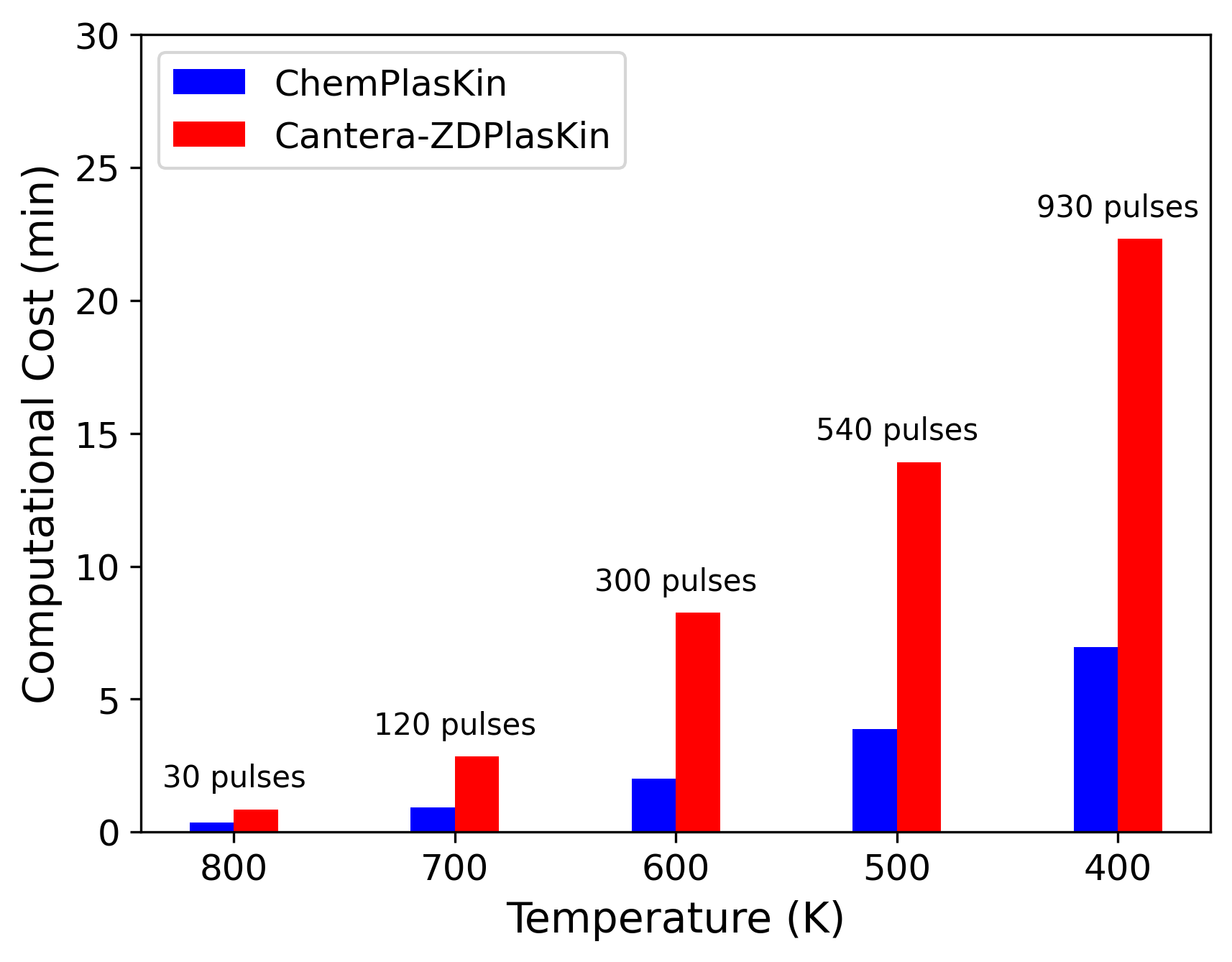}
\caption{\label{fig:compu_cost} Computational cost (clock time) comparison between ChemPlasKin and Cantera-ZDPlasKin corresponding to the NRP assisted ignition depicted in Figure \ref{fig:IDT}. Single-thread execution at 2.3 GHz. Profiling results confirm that there is no data communication bottleneck in the Cantera-ZDPlasKin solver.}
\end{figure}

A comparison of computational efficiency between ChemPlasKin and Cantera-ZDPlasKin was carried out for NRP discharge assisted ignition cases, as shown in Figure \ref{fig:compu_cost}. ChemPlasKin achieved a three-fold speed-up. Note that in Figure \ref{fig:compu_cost}, ChemPlasKin employed the same dynamic timestep settings in the main loop to ensure a fair comparison. In actual applications, ChemPlasKin can utilize much larger outer timesteps, as it is not constrained by operator-splitting errors.

Note that many operator-splitting solvers listed in Table \ref{tab:1}, which operate based on kinetic mechanisms developed in various studies, lack rigorous validation against experimental data at micro time scales, such as radical production and fast and slow gas heating during a single NRP discharge period. The following two subsections will present validation of ChemPlasKin that addresses this gap, similar to the approach taken by Cheng et al. \cite{cheng_plasma_2022}.

\subsection{Spark discharge in air} \label{spark_air}
\begin{figure}[H]
\centering
\includegraphics[width=0.6\linewidth]{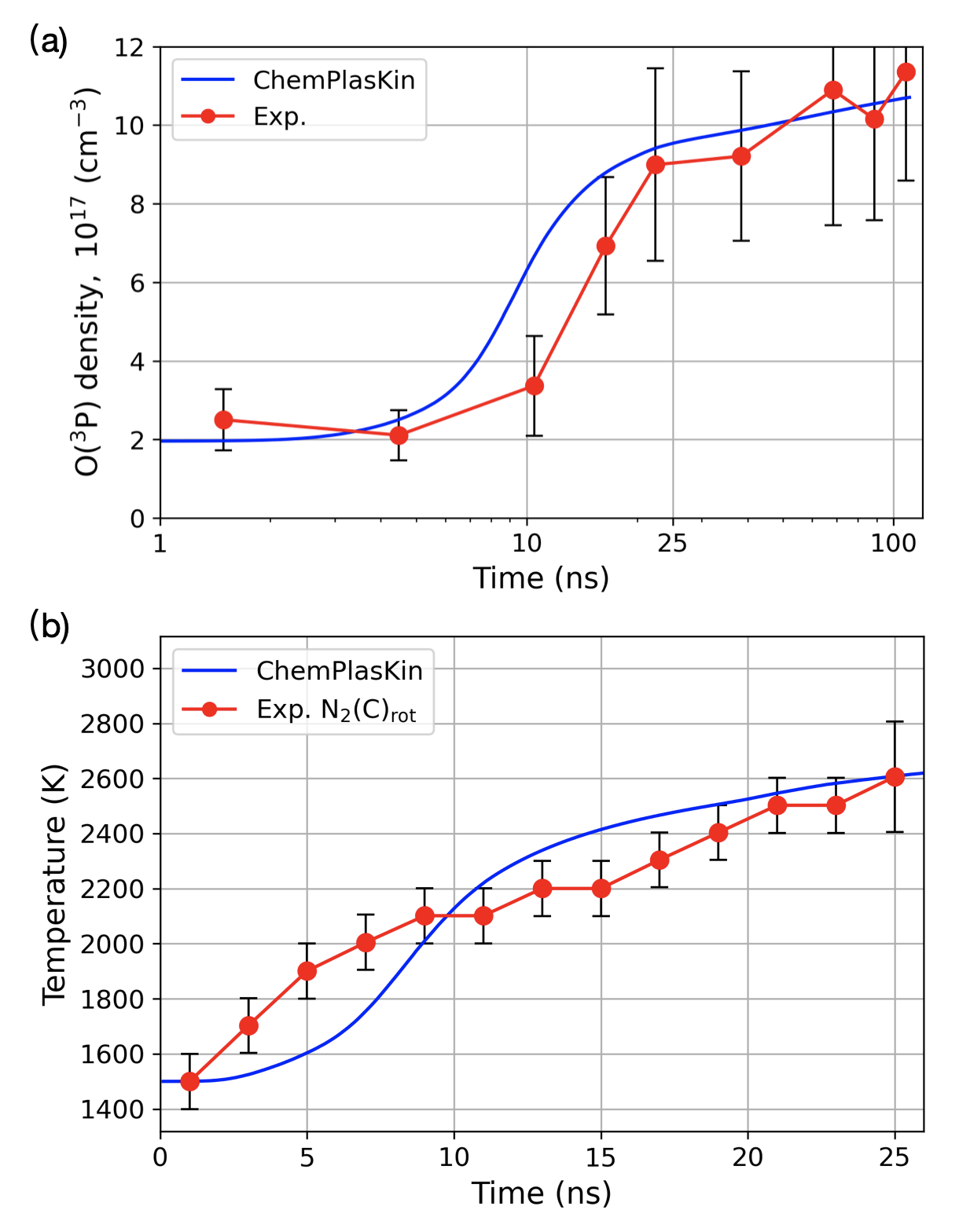}
\caption{\label{fig:spark} Comparison of ChemPlasKin predictions with experimental data on the temporal evolution of (a) $\mathrm{O(^3P)}$ production during and after a nanosecond pulse and (b) gas temperature, demonstrating good agreement. Experimental data from \cite{rusterholtz_ultrafast_2013}.}
\end{figure}

This test case aligns with Case A from \cite{cheng_plasma_2022}, utilizing the experimental setup described in \cite{rusterholtz_ultrafast_2013}, where NRP spark discharges are generated between pin-pin electrodes. The gas temperature can be inferred from the rotational temperature of $\mathrm{N_2(C)}$ for a duration of at least 17 ns \cite{rusterholtz_ultrafast_2013}. The initial gas mixture temperature is set at 1500 K with a composition of $\mathrm{N_2:O_2:O = 77.4:18.6:4}$, accounting for the thermal and chemical effects of preceding pulses. Figure \ref{fig:spark} presents a comparison of the gas temperature evolution and $\mathrm{O(^3P)}$ production as predicted by ChemPlasKin against experimental data.

\subsection{V-T relaxation in air} \label{v-t}
\begin{figure}[H]
\centering
\includegraphics[width=0.6\linewidth]{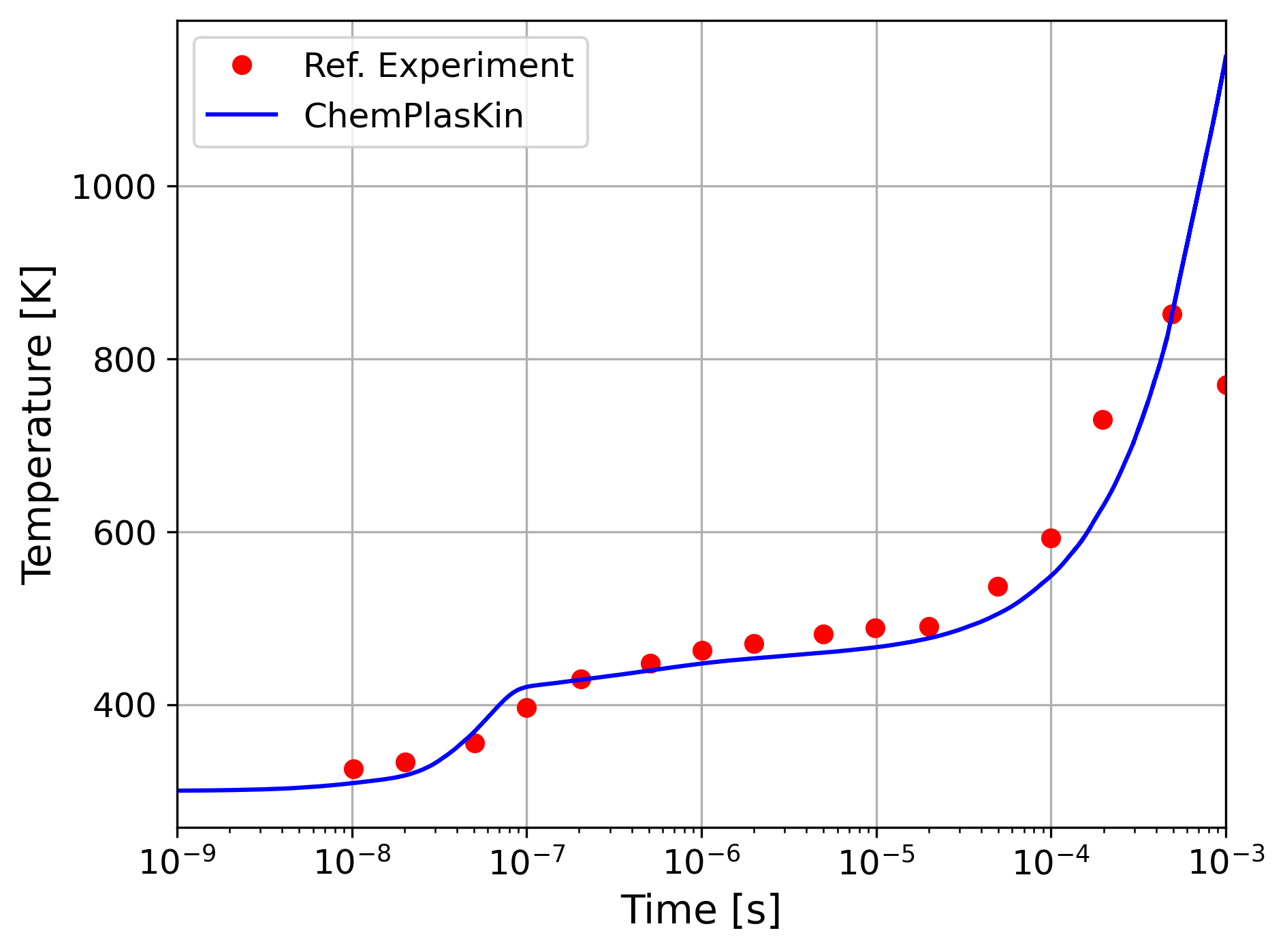}
\caption{\label{fig:T_VT} ChemPlasKin predictions of gas temperature from VT-relaxation of $\mathrm{N_2}(v)$ in pure air, showcasing a comparison against experimental measurements from \cite{montello_picosecond_2013}.}
\end{figure}

This validation case aligns with Case C from \cite{cheng_plasma_2022}, utilizing the experimental setup detailed in \cite{montello_picosecond_2013}. The initial conditions of air are set at a temperature of 300 K and a pressure of 100 Torr. Between the pin-pin electrodes, the applied $E/N$ is approximately 100 Td, with up to $50\%$ of the plasma energy contributing to the vibrational excitation of $\mathrm{N_2}$. In this scenario, the model's prediction of the rotational temperature, represented by the solid curve shows good agreement with the experimental data, as shown in Figure \ref{fig:T_VT}. Note that the simulation focuses solely on $\mathrm{N_2}$ for the vibrational energy equation, under the assumption of isobaric conditions due to the minimal impact of fast gas heating in this context. The exclusion of heat loss from thermal diffusion in the model accounts for the observed discrepancies between calculated results and experimental measurements on the millisecond timescale.

\subsection{Hydrogen oxidation with DBD discharges} \label{dbd}
This validation case compares ChemPlasKin's performance against both experimental and numerical results detailed in \cite{snoeckx_kinetic_2022}, which investigates the kinetics of plasma-assisted oxidation of $\mathrm{H_2}$ in an undiluted lean $\mathrm{H_2/O_2}$ system. In the referenced study, a preheated gas mixture is subjected to treatment in a dielectric barrier discharge (DBD) reactor under constant pressure and temperature conditions. To approximate the filamentary nature of the discharges, it is assumed that each fluid particle encounters a total of 288 rectangular pulses at a constant $E/N$. 

\begin{figure}[H]
\centering
\includegraphics[width=0.6\linewidth]{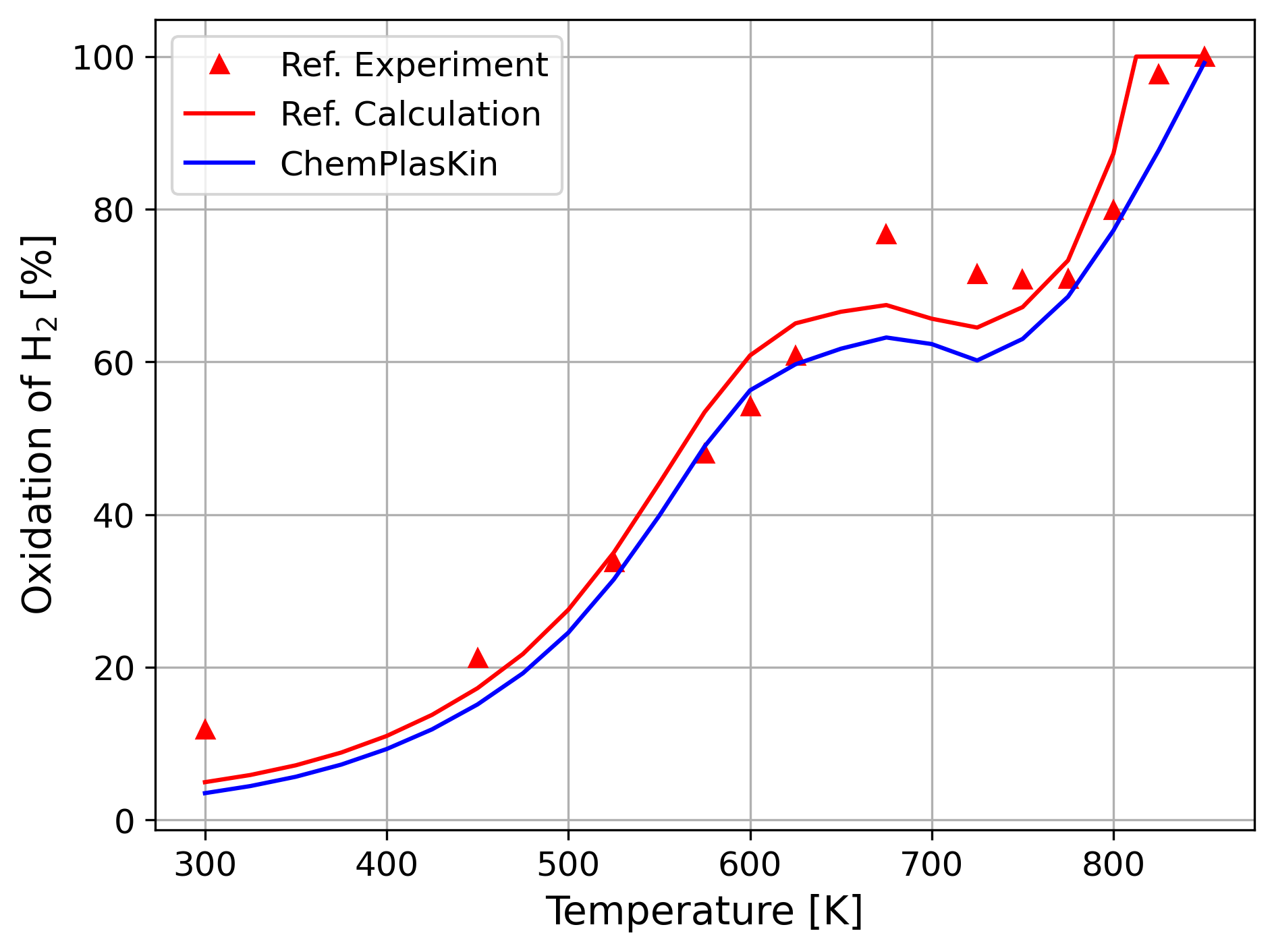}
\caption{\label{fig:h2o2} Comparison of ChemPlasKin simulations with reference data on plasma-assisted oxidation rates of $\mathrm{H_2}$ over a range of gas temperatures ($T_{\text{gas}}$) for an $\mathrm{H_2/O_2}$ mixture ($\phi=0.01$) at 10 W of plasma power. Reference data from \cite{snoeckx_kinetic_2022}.}
\end{figure}

Figure \ref{fig:h2o2} presents a comparison of the $\mathrm{H_2}$ oxidation rates derived from ChemPlasKin against the experimental and numerical results under varying gas temperatures.
The agreement between ChemPlasKin's predictions and the reference results is generally good, with a larger discrepancy at higher temperatures, potentially attributed to the omission of thermal cracking residence time in our model. ChemPlasKin completes each data point presented in Figure \ref{fig:h2o2} within approximately 45 seconds of single-threaded execution on a 2.3 GHz core.

\subsection{Heat loss model} \label{heat_loss}
\begin{figure}[htbp]
\centering 
\includegraphics[width=0.6\linewidth]{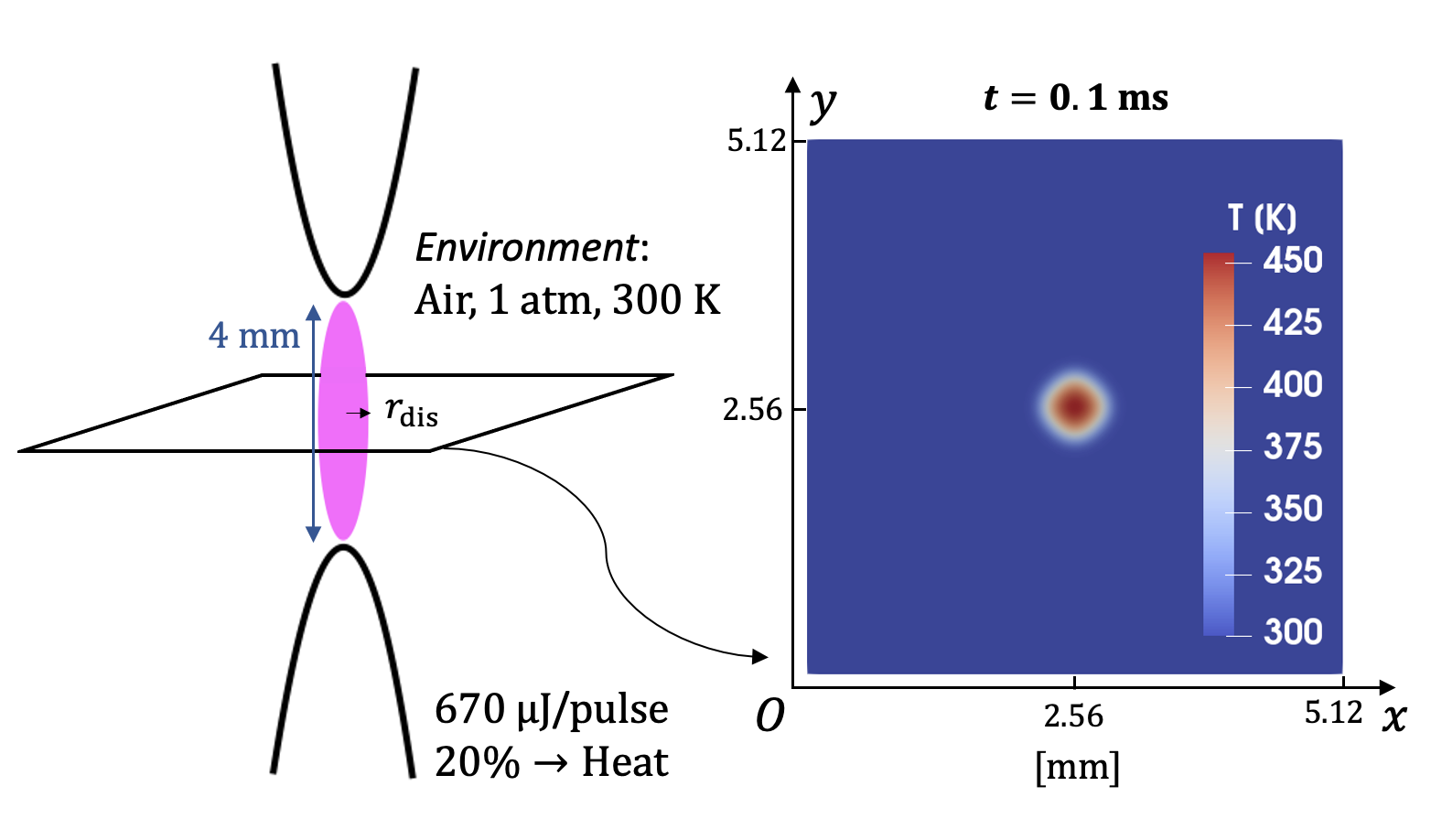}
\caption{\label{fig:2D-config} Schematic of a 2D computational domain for simulating ultrafast heating effects from pin-pin generated NRP discharges using phenomenological model.}
\end{figure}

The proposed heat loss model, detailed in Section \ref{heat_loss}, is validated focusing on the fast gas-heating phenomenon, thereby omitting the need for detailed plasma kinetics and slow gas heating. Figure \ref{fig:2D-config} illustrates a typical scenario for applying the heat loss model, aligned with the test case described in Section \ref{spark_air}. Here, NRP discharges at 10 kHz are generated between pin-pin electrodes, with each pulse delivering an energy input of $670\ \mathrm{\mu J}$. According to the phenomenological plasma model by \cite{castela_direct_2016}, $20 \%$ of the pulse energy is allocated to fast gas-heating within a 10 ns duration. A horizontal 2D computational domain is defined, featuring the pulsed heating source at its center, modeled with an energy density spatial function $\mathcal{F}=\text{erfc} \left(r^2/{6.9\times 10^{-8}} \right)^{2.5}$, where $r$ denotes the radial distance from the discharge center and an effective discharge radius $r_\text{dis}$ of $225 \ \mathrm{\mu m}$ is specified. The domain employs a non-reflective boundary condition, ensuring the pressure field near the discharge zone recovers the ambient level within microseconds \cite{shao_computational_2024}.

\begin{figure}[htbp]
\centering 
\includegraphics[width=0.6\linewidth]{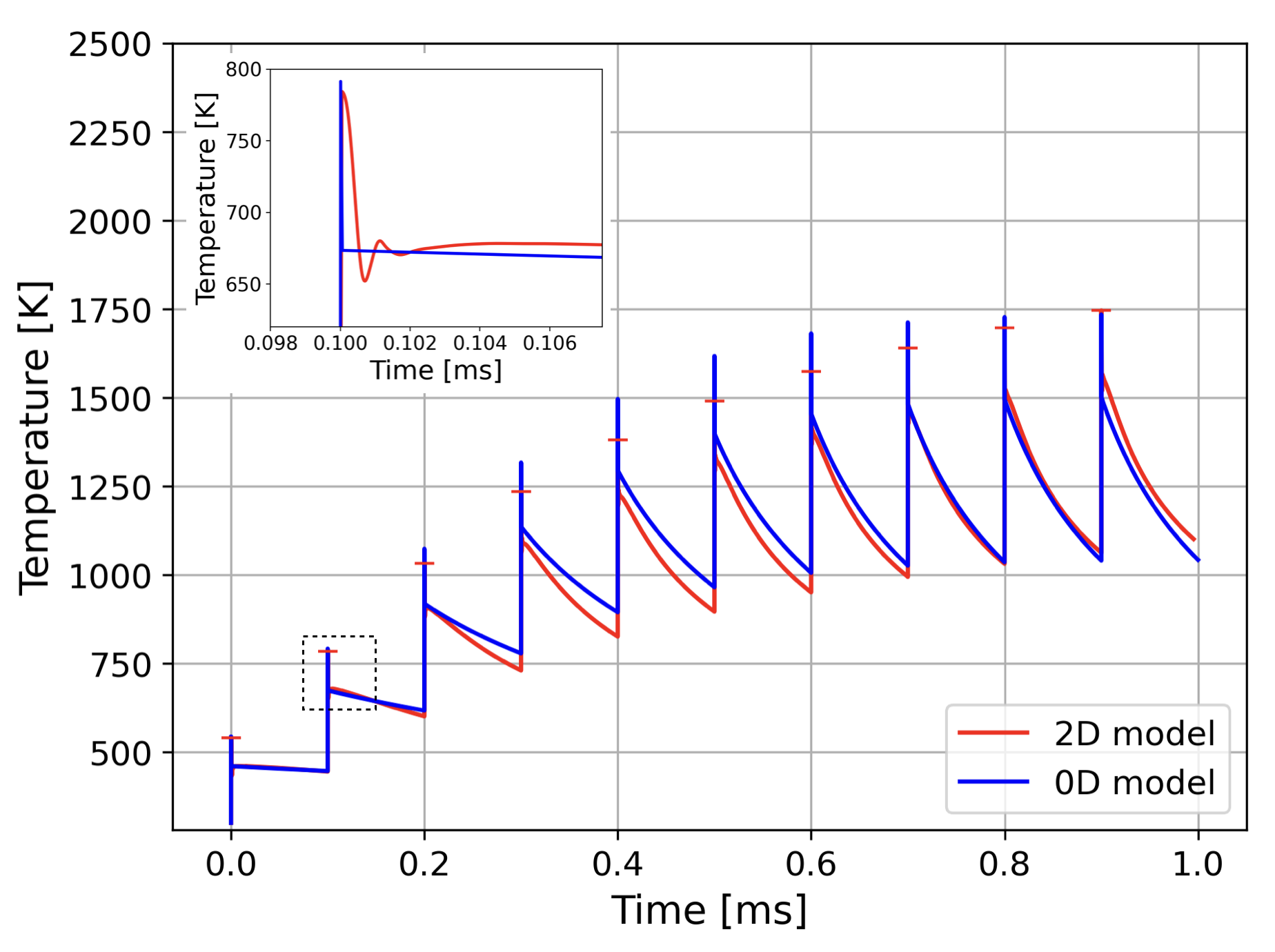}
\caption{\label{fig:heat_loss} Comparison of temperature evolution within the discharge kernel across 10 NRP discharges between 2D and 0D phenomenological models. V-T relaxation heating not included. The red bars highlight the highest temperature for each nanosecond pulse in the 2D model. $C_0=1.0$ is employed in the 0D model.}
\end{figure}

The averaged temperature within $r_\text{dis}$ in the 2D domain, solved using OpenFOAM \cite{weller_tensorial_1998}, represents the discharge kernel temperature. In parallel, a 0D model, simply written as an independent Python code, utilizes the Cantera library to update thermal conductivity based on temperature changes. After applying ten heating pulses, Figure \ref{fig:heat_loss} displays the temperature comparison between the 2D and 0D models. The models demonstrate good overall agreement, with the 0D model's assumption of instantaneous isentropic expansion capturing the sharp temperature drop at each pulse's conclusion, as elaborated in the zoom-in subplot of Figure \ref{fig:heat_loss}. Without incorporating the heat loss model, a conventional 0D chemistry-plasma model would likely predict a step-wise, monotonic temperature increase for NRP discharges, contrasting with the quasi-steady state indicated by the 2D model.

\begin{figure}[H]
\centering
\includegraphics[width=0.6\linewidth]{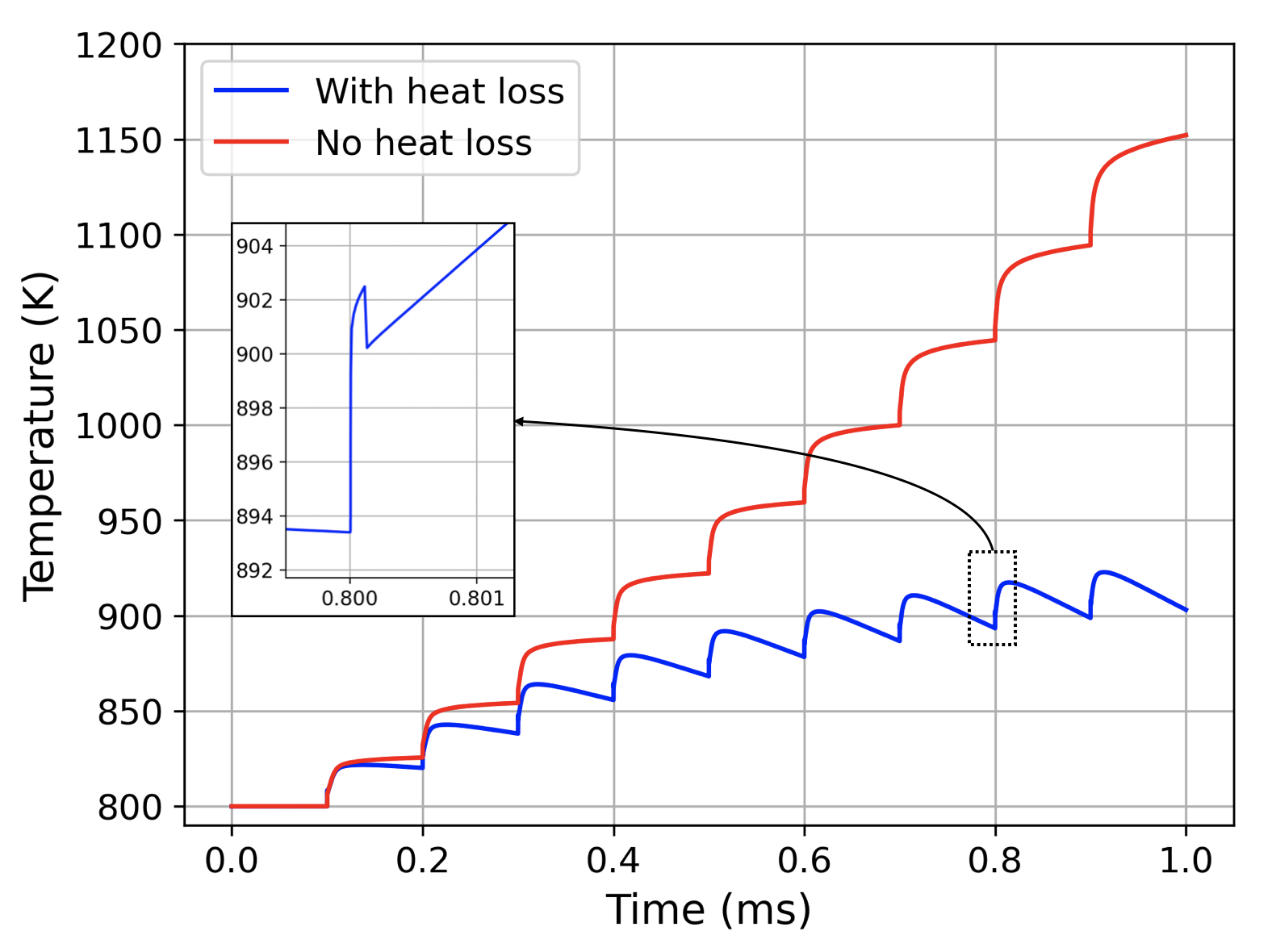}
\caption{\label{fig:heat_loss_detail} Temperature evolution in a $\mathrm{0.0499 CH_4/0.7505 N_2/0.1996 O_2}$ mixture with NRP discharge assisted ignition over the first ten pulses. Deposited plasma energy is $\mathrm{10\ mJ/cm^3}$ per pulse, with $C_0=1.0$.}
\end{figure}

Following the standalone examination discussed previously, the impact of the heat loss model on gas temperature in ChemPlasKin, which incorporates detailed gas-plasma kinetics of a methane/air mixture \cite{cheng_plasma_2022}, is presented in Figure \ref{fig:heat_loss_detail}. The discrepancy in temperature evolution with and without the heat loss model during the first ten pulses clearly demonstrates its potential influence on ignition delay time calculations. We acknowledge that further validation and improvement of this model are necessary, ideally through direct comparison with experimental data or high-fidelity multi-dimensional simulations.

\section{Conclusions}

This study describes ChemPlasKin, a freeware tool developed for simulating gas-plasma kinetic processes. ChemPlasKin integrates the electron Boltzmann equation solver CppBOLOS with the Cantera library, enabling the solution of neutral gas and plasma kinetics within a unified ODE system. Additionally, a supplementary heat loss model is proposed to enhance the accuracy of temperature predictions for nanosecond repetitively pulsed (NRP) discharges in configurations utilizing pin-pin electrodes.

To evaluate the computational efficiency of ChemPlasKin, we constructed a Cantera-ZDPlasKin PAC solver using the widely used operator-splitting method. In the test cases, this configuration achieved a threefold speedup, and we anticipate even faster performance with a relaxed outer timestep.

The C++ solver has been validated against experimental results across various aspects and timescales, including ultrafast gas heating and radical production, slow gas heating from V-T relaxation and fuel reforming involving hundreds of pulses. ChemPlasKin shows its capability as a versatile tool, particularly useful for PAC and fuel reforming. It can also be used purely as a plasma kinetics solver, akin to ZDPlasKin. Its development provides a resource for researchers seeking to explore gas-plasma kinetics without the need to invest in custom coding efforts. Continued validation and enhancement by the user community are encouraged. We also anticipate ChemPlasKin to be incorporated into CFD codes to enable high-fidelity simulations of fully-coupled plasma-assisted reacting flows, acknowledging that computational costs remain a significant consideration.

\section*{Acknowledgments}
This research was funded by King Abdullah University of Science and Technology (KAUST). The authors would like to thank Seunghwan Bang and Dr. Ramses Snoeckx for useful discussion. The first author also thanks friends Renston, Vijay, and Alessandro for the great tunes and good times that helped lighten the load.

\bibliographystyle{unsrt}
\bibliography{My_Library}

\end{document}